# Understanding Software Developers' Approach towards Implementing Data Minimization

Awanthika Senarath
School of Engineering and IT
University of New South Wales
a.senarath@student.unsw.edu.au

Nalin A. G Aarachchilage
School of Engineering and IT
University of New South Wales
nalin.asanka@adfa.edu.au

## ABSTRACT
Data Minimization (DM) is a privacy practice that requires minimizing the use of user data in software systems. However, continuous privacy incidents that compromise user data suggest that the requirements of DM are not adequately implemented in software systems. Therefore, it is important that we understand the problems faced by software developers when they attempt to implement DM in software systems. In this study we investigate how 24 software developers implement DM in a software system design when they are asked to. Our findings revealed that developers find it difficult to implement DM when they are not aware of the potential of data they could collect at the design phase of systems. Furthermore, developers were inconsistent in how they implemented DM in their software designs.

## 1. INTRODUCTION
Data Minimization (DM) is a simple and a straight forward privacy practice which instructs the minimization of the use of personal data in software systems [7]. DM requires software systems to not use user data for purposes other than for which it was collected, and to avoid collecting data that is not absolutely necessary for the purpose of the system [19]. However, in a recent privacy incident at Facebook, Cambridge Analytica, a data analytics company, was able to access data of 50 million Facebook users for a purpose which was not known to the user [2]. This happened because Facebook did not have DM embedded into its design.

Large Scale Data Analysis (LSDA) and surveillance systems (Internet of Things) that continuously collect user data are the biggest challenges of DM [16]. The potential of these systems to utilize data to monitor user behaviors that add value to businesses (such as marketing by identifying customers who are likely to buy certain products) discourages software system developers to consider DM when they design systems [16, 14]. At the same time, the difficulty to relate DM requirements into software implementation techniques, such as anonymization and encryption, also makes it difficult for software developers to successfully implement DM in their software system designs [15, 12]. However, software developers disregarding DM requirements when they design software systems could result in privacy invasions similar to the Cambridge Analytica scandal in Facebook, which caused a loss of USD 70 billion or more for Facebook [2]. Especially with the new European Union Data Protection Regulations (GDPR) reforms enforcing explicit consent from users for data collection for clearly defined purposes [19], developers will be required to justify the reasons and get user consent when using user data in software systems. Therefore, it is important to understand how software developers implement DM in software systems, and the problems they face. This would enable developing privacy guidelines for software developers that solve the real problems they have.

In order to understand how software developers implement DM in software systems, in this study we recruited 24 software developers via Github and assigned a simple software system design task to them. We provided them with a specification for a software system. We hinted on the potential of using LSDA in this system in order to make it a typical large scale data collection system. We then explicitly instructed developers to practice DM in their designs, because we were interested in conscious behaviour. Through their answers to a short exit questionnaire, we investigated how developers implement DM in a software system that has a potential for LSDA. Our findings revealed that developers had difficulty in limiting data collection in the software system as they were unable to pre-determine the potential of the data they could collect. Because of this when developers embedded DM into their designs, they tried collecting more data and controlling the use of data in the system. These findings could help privacy researchers and software development organizations to understand the practical problems developers have in implementing DM in software system designs, and to guide developers effectively to solve the problems they face.

## 2. RELATED WORK
Introducing different ways to implement DM in software systems has been a focus of privacy researchers. For example, Pfitzmann et al. instruct implementing DM with un-linkability, un-observability, and pseudonymity, where all of them are techniques on collected data within software systems [13]. Gurses [10] suggests implementing DM as a binary concept, where disclosing data leads to loss of privacy and non disclosure leads to privacy. Makker et al. [11] propose a DM scheme with edge processing, where only aggregated data is sent into processing. In another approach, Agrawal et al. [4] propose hypocratic databases, a data minimization scheme focusing on data storage in databases. All these approaches attempt to provide instructions on how developers should implement DM in their system designs. However, it is said that when implementing DM in software systems, developers are always in a dilemma to create more features in the system with more data, and to ensure user privacy by not using too much data [6]. How developers attempt to solve the problems they face when they design software systems are not known to us.

For example, it is said that developers tend to ignore privacy instructions that deviate from their usual software development frameworks [5]. Furthermore, it is said that when system requirements contradict with privacy requirements, developers prioritize system requirements over privacy [15]. Therefore, observing how actual developers implement DM and the problems they face would help us to understand why developers still fail to implement DM effectively in their system designs. Furthermore, this knowledge could be used to effectively guide developers to implement DM in system designs and address the problems they face.

## 3. METHODOLOGY

Our goal in this research is to investigate how software developers implement DM when they design software systems. We conducted the study remotely, so that developers could participate from their workplace or home, therefore behaving closer to how they really work. We recruited software developers for the study through GitHub [3]. We sent invitations to publicly available email addressed in GitHub for active committers in a selected set of most active Java and PhP projects. We sent them information about the study and ourselves in the invitation email and asked them to reply to express their interest for participation. We sent 6000 emails and received 118 expressions of interest. For those who were interested in participation, we sent a second email with the ethic consent form and study instruction form with details on how to participate in the study. After sending the second email, we received 25 answers, out of which one participant's answers were rejected due to lack of quality. Each participant spent an average of 2.5 hours in the study. The complete study design was approved by the university ethic committee.

We gave the participants an application scenario and asked them to practice DM in the design. The scenario we gave is as follows,

*System Specification given in the design task :* A web-based health-care application that allows remote consultation with medical professionals, general practitioners and specialists, for a payment. Users should be able to browse through a registered list of medical professionals and chat (text/video) with them on their health problems for advice. Doctors and health-care professionals can register on the application to earn by providing their expertise to users. The application is to be freely available on-line (desktop/mobile). You may consider sharing user data with hospitals, insurance companies and advertisers to gain profit.

We then asked the participants to decide the user data (such as name, address) they want to collect for the application. Then we asked the participants to draw the information flow diagram for the application to see how they make decisions on data storage and sharing considering the DM concept in the design. At the end of the task we asked them,

- Did you use DM in designing this software application?
- Please elaborate on how you used it, if you did not use the concept, please give your reasons.

By asking them what they did in the task they just completed, we attempt to eliminate limitations in the answers due to participants' recalling capacity and memorability [3]. With this information we aimed to observe how a developer would practice DM in a system design and the issues they had in implementing DM in a system design.

We qualitatively analyzed the answers to open ended question to explore the reasons for the yes/no answers. We used two coders to code the answers in Nvivo [1] using the grounded theory approach, where coders coded data and tried to extract information from the answers without prejudice [9]. Both coders had 9 similar codes, with 6 codes being present in either one of the coders. Then the coders discussed between themselves and resolved the disputes by summarizing the codes iteratively. They finally came up with 10 final codes under two themes. From these codes we removed codes that were related to organizational policies and management support as this was out of the scope of our research. In this work we are only interested in how developers approach DM when they design software systems and their expertise on DM. We discuss these in detail in the results section.

## 4. RESULTS AND DISCUSSION

20/24 participants said yes when they were asked if they used DM in their system design. From this result one can assume that, when developers are instructed to use DM in a system design task, majority of them use it in their design or rather claim that they have used it in their design. However, when we analyzed the descriptive answers we observed that 8 of the participants who claimed they used DM in the design said their implementation of DM may not be effective and that they were not sure if they implemented DM correctly. For example, P12 said *I used data minimization, but I don't think I have made a good one, I tried to make it, but I found myself still using much private data of users to keep people involved real*. P23 said *I did, but not in the most effective way*. This suggests that the actual number of developers who really used DM in their system design could be much less. Therefore, just because a developer claims s/he has used DM in a system design, it does not guarantee that the system has the requirements of DM in its design. Developers gave several reasons as to why they could not use DM effectively in their design. Figure 01 shows codes related to the concerns developers had.

All developers who did not use DM said they lacked experience using DM, and that they did not know the concept (4/24). Some of those who said they used DM were not sure if they used it in the design effectively (8/24). Developers were not confident about the way they used it because they did not get any feedback. P9 said *formalise these processes so that we \*know\* it is being done right*, suggesting developers require systematic guidance that give them feedback on what they are doing. At the same time, developers were concerned about the time and effort required to implement DM in the system design. P9 said *considering all these things makes projects MUCH more involved than originally expected*. When the effort required for implementing DM in the system was high developers were likely to give up.

---
[1]https://www.qsrinternational.com/nvivo/

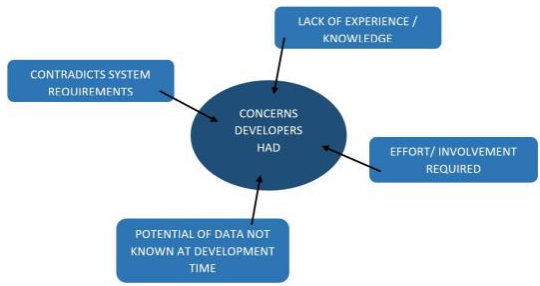

Figure 1: Developers concerns in implementing DM

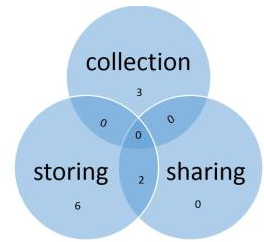

Figure 2: Areas of focus when implementing DM

When developers tried using DM in the design, they said they had concerns in minimizing the collection of data for the system (19/24) as they did not know the full potential of data at the development time. Participants who considered DM to be only about collection limitation said they could not effectively implement it. For example, P8 said *the requirement for the application makes it hard to predict the usefulness of the data* and P24 said *sometimes that information can be used for automatic analysis*.

The inappropriateness of the concept of DM in controlling only the collection of data has been previously raised by researchers focusing on LSDA techniques [7]. For example, Tene and Polonetsky [17] emphasize that with the context of big data, the principle of DM should be interpreted differently, requiring organizations to de-identify data when possible and implement reasonable security measures because big data business model in itself is antithetical to the current definition of DM. In developers' efforts to apply DM while balancing system functionalities and benefits user data could deliver, they ended up focusing more on limiting data storage compared to data collection. Therefore, developers considered data minimization to be a concept that goes beyond the data collection phase as they were making an effort to practically implement DM in their system designs, without compromising the benefits of LSDA. For example, P6 said *I limited it [DM] to storing only relevant data, and sharing the minimum to each party that requires it. my design tries to satisfy the project requirements while keeping the data minimization, but they come often contradictory*. Figure 2 shows the different phases of data usage in the system developers focused on when implementing DM in their system design.

Most of those who implemented DM focused on minimizing data storage rather than minimizing data collection or sharing. Two participants focused on both sharing and storage. None of the participants focused on all three phases of data usage in the system. However, it has been argued that DM is not only about what data points you collect, but the associations you make amongst the data [1]. Therefore, a comprehensive DM implementation may have to focus on the entire data processing chain (collection, recording, storage, alteration, linking and consultation [18]) in a system. Unfortunately, none of the developers recognized this need.

We also observed that different developers used different privacy techniques to implement DM in their system design. Figure 3 summarizes the ways in which developers implemented DM in their system designs.

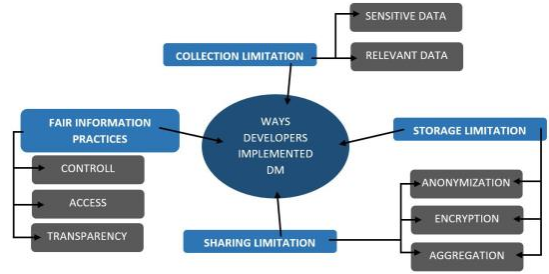

Figure 3: How Developers implemented DM in their designs, first level of rectangles are the summarized codes, and the second level of rectangles shows the exact techniques used by developers

Interestingly some participants also implemented principles in the Fair Information Practices (FIP) [8], such as access and control, in their efforts to implement DM in the system design. P18 said s/he made data entry fields optional and gave control to the user to expose data in the system. P24 said s/he gave information on using and collecting data to the user in order to ensure DM is implemented in the system.

Some developers used techniques such as aggregation, encryption and anonymization to implement DM in the design. Two participants identified anonymization could be used to implement DM and one identified aggregation could be used. They suggested that if the system uses anonymized data this essentially satisfies the DM requirements. One participant suggested using a pseudonym to minimize using user data in the system *patients name is not required. Instead we can come up with patient id*. Two of the participants implemented encryption in the design, saying if data is encrypted in a way that it can only be accessed and used by the user, data is not used in the system. However, the percentage of developers who identified the implementation techniques for DM in the system design was not satisfactory (8/24). Most developers could not understand how DM could be implemented in the system (15/24). For example, P10 said *a*

*developer just needs to get stuff out the door so they can eat. Although I take privacy and security very importantly, it as mostly, in my experience, focused on the technology to support this*. Developers preferred technical instructions than concepts to guide them. Previous research has claimed that developers fail to identify that techniques like anonymization could be used to implement DM in a system [12]. This is because DM only requires developers to minimize the use of data, and does not specify how developers should do it. The lack of explicit instructions, therefore, frustrates developers. For example, one participant had concerns about using anonymization to implement DM, as it could hinder the benefits of data analysis. S/he said, *If the insurance policy company wants to correlate Polish immigrants with diseases, they need last names or ethnic backgrounds. Would the anonymization process retain enough of a correlation there?*. Therefore, if developers are guided with explicit instructions on how to minimize the use of data while achieving system requirements, developers would be able to implement DM in their system designs better.

## 5. CONCLUSIONS

This study investigates how software developers implement DM is software system designs. Our main findings are,

- Developers have difficulties in satisfying DM requirements when they could not pre-determine the benefits LSDA could bring into the system.
- Developers tried to expand the principle of DM across the complete data processing chain within the application (collection, storage and sharing) to minimize using user data in the system.
- Developers were inconsistent in the areas they focused (collection, storage) and the techniques they used (encryption, aggregation) to implement DM in their system designs.

These findings could help researchers, organizations and privacy experts to establish better guidelines to encourage developers to practice DM in LSDA systems successfully. As future work we aim to construct a novel approach for DM that address the problems developers have, in order to enable minimizing the use of personal data in software systems without compromising system functionalities.